# Experimental demonstration of novel optical Brownian ratchet by controllable phase profile of light

Xionggui Tang*, Yanhua Xu

*Department of Physics, Key Laboratory of Low Dimensional Quantum Structures and Quantum Control of Ministry of Education, Synergetic Innovation Center for Quantum Effects and Applications, Hunan Normal University, Changsha, 410081, P.R. China*
*\* tangxg@hunnu.edu.cn*

**Abstract**: Brownian ratchet has emerged as a promising tool for understanding motion mechanism of molecules and proteins, and dynamically manipulating particles in non-equilibrium thermodynamics state. Here, we propose and experimentally demonstrate a new type of optical Brownian ratchets, which is generated by using controllable phase profiles in holographic optical trapping system. The potential energy profiles are dynamically switched in on-off mode, controlled by Labview software. Experimental results reveal that not only high speed but also large step distance can be easily achieved in Brownian ratchet, in which the average velocity of forward motion is about 28 μm/s and step distance is about 42 μm. More importantly, its motion path, step distance and velocity can be easily changed by manipulating its holograms, which exhibits high flexibility, easy control and excellent capability. This technique provides new opportunities for exploring non-equilibrium dynamics at the nanoscale level, and developing novel nanoparticle manipulation.

Typically, micro- and nano-particles in various solutions are frequently collided by solvent molecules, leading to Brownian motion [1]. This motion can be rectified by breaking its spatiotemporal symmetry, and then a net directional motion appears, called as Brownian ratchet motion [2-3]. This fascinating phenomenon, essentially originating from the nonequilibrium thermodynamics at submicron scales, has attracted rapidly growing interest in different scientific areas, from molecules motor, protein translation to transport and separation of particles [4-5]. In past two decades, several proposals have been demonstrated for realization of different ratchet motions [6-21]. Among them, optical Brownian ratchets have become one of the most promising manipulation systems in controllable directional transport of particles. In 1995, L. P. Faucheux et. al. proposed an approach for optical Brownian ratchet, realized by dynamically scanning the beam intensity along circle orbit in a spatially periodic but asymmetric way [12]. In 2005, Jaehoon Bang et. al. experimentally demonstrated an optical Brownian ratchet using multiple spot optical tweezer modulated by acousto-optic deflector, to spatially generate flat and saw-tooth trapping potentials, respectively [22]. In 2016, Shao-Hua Wu et. al. presented optical Brownian ratchet by employing near-field on-chip traps on an asymmetrically patterned photonic crystal [3], but its design and fabrication which is required to be highly accurate at nanometer scale, remains very difficult. Actually, the approaches mentioned above are realized by using intensity gradient force. The major disadvantages of particle transport are obvious, including low speed, small step distance, the limited motion path and the complicated experimental system. Specially, the volume of particle is required to be large, due to slower Brownian motion and larger scattering cross section. In contrast, the optical ratchets based on phase gradient force have not been reported yet.

In this work, we report an optical Brownian ratchet using controllable phase profiles in holographic optical trapping system. The proposed optical ratchet has remarkable advantages such as flexible design, easy control, high motion speed, large step distance and variable motion path. The experimental results show that the average velocity of above 28 μm/s can be easily achieved in nanoparticle manipulation, which is much greater than previously reported ones. More importantly, the average velocity can be easily manipulated by changing its phase profile.

The physical principle of Brownian ratchet is described in Fig. 1. For simplicity, the particles are assumed to diffuse in a one dimension. A spatially asymmetric potential energy profile and a flat potential energy profile are dynamically switched in an on-off mode, in which duration time are $T_1$, $T_2$ and $T_3$ in one cycle, respectively. In first stage (illustrated in I), the particles are trapped in potential wells, and no net directional motion appears, in which its potential energy profile is shown in green, and the probability distribution is given in blue (top panel). At second stage (illustrated in II), its potential energy profile is switched into a flat profile, and then particles diffuse freely in a way of Brownian motion, in which its probability is described by broadened Gaussian distribution (middle panel). At third stage (illustrated in III), asymmetric potential energy profile is turned on again. At this moment, if the particle stays in right region of vertical dashed line (the probability is shown in red), it will move forward to the right adjacent trap (bottom panel), leading to a net directional motion. Otherwise, it is trapped in original wells. In addition, the asymmetry of potential energy profile is characterized by the parameter α, calculated by $l_1/(l_1+l_2)$, in which $l_1$ and $l_2$ stand for the forward and backward length, and the former is usually larger than the latter. The probability of forward motion in one cycle can be varied by changing the parameter α.

How can we experimentally realize such dynamically modulated potential energy profiles? Previously, the single optical spot have been usually employed in dynamically scanning way to generate asymmetrical and flat potential energy profiles, whose disadvantages such as low speed and small step distance limit its manipulation capability. Notably, optical force can be created by both intensity and phase gradient profile [23-31], and it can be written as,

$$\bar{F} = \frac{1}{4}\varepsilon_0\varepsilon\alpha'\nabla I + \frac{1}{2}\varepsilon_0\varepsilon\alpha''I\nabla\varphi, \qquad (1)$$

where $\varepsilon_0$ and $\varepsilon$ are dielectric constant and relative dielectric constant, $\alpha'$ and $\alpha''$ are the real and imaginary part of particle's polarizability, $I$ is intensity distribution, and $\varphi$ is the transverse phase profile of light. If the intensity is uniform in the operation region, the intensity gradient force will be zero. Therefore, optical driving force is only induced by phase gradient. Thus, its potential energy profiles can be easily obtained by,

$$V = -k\varphi, \qquad (2)$$

where $k = \varepsilon_0 \varepsilon \alpha'' I / 2$. It reveals that there is anti-symmetrical relationship between potential energy profiles and phase profiles. Herein, we demonstrate a novel approach for realizing optical Brownian ratchet by using dynamically controllable phase profiles. To our knowledge, asymmetrical potential energy profiles are firstly sculpted by using phase gradient profile. In our experiment, holographic optical tweezers (HOTs) are employed for spatially generating potential energy profiles, and are dynamically modulated by Labview software. The phase profiles are depicted in black line (illustrated in IV). Its phase profile is asymmetrical at first and third stage, but it are uniform at second stage. Importantly, all of intensity profiles are uniform (not illustrated here). Their potential energy profiles can be dynamically switched from one to another while changing its phase profiles, which has promising advantages, including flexible design, easy control and excellent manipulation performance in optical Brownian ratchet. While the intensity gradient is employed to create such asymmetrical potential profiles, however, it still remains challenging to accurately realize asymmetrical potential wells in such a large operation area.

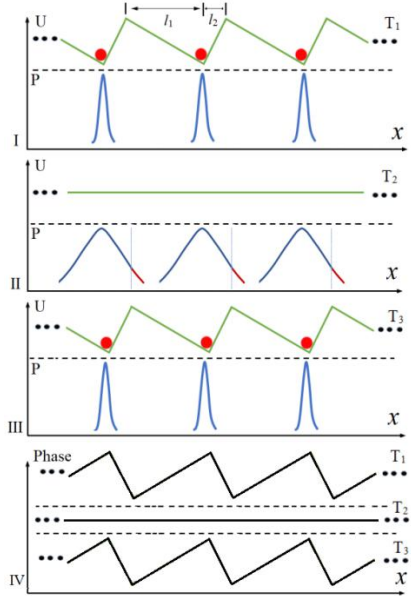

**FIG. 1.** Scheme for optical Brownian ratchet. Panels I-III correspond to potential energy profiles and probability at time interval $T_1$, $T_2$ and $T_3$, respectively, and Panel IV denotes its phase profile at corresponding time.

In our experiment, ring-shaped trap is chosen as an example for demonstrating optical Brownian ratchet. Firstly, we design three different holograms (see hologram design and Fig. S1 in Supplementary Material). The first hologram is used for generating flat intensity and phase distribution along the circle orbit, and the other two holograms are designed to create flat intensity and asymmetrical phase gradient profile. The corresponding phase profiles are shown in Fig. 2(a), whose ring radius is about 7 μm. It reveals that reconstructed phase profiles are in good agreement with our target profiles, which further demonstrates that our designed holograms are very accurate. Specially, the difference between them in Fig. 2(a) (II and III) is that step distance and the number of asymmetrical potential wells are different. In addition, their simulated and measured intensity distributions are provided (see Fig. S1). It finds that all of intensity distributions are nearly uniform along circle orbit, which well meets our design requirement. In the following, we calculate their potential energy profiles, as given in Fig. 2(b). Here, it needs to mention that the potential energy induced by intensity gradient have not been included in ring traps with asymmetrical potential energy profiles, because their intensity distributions are almost identical along the circle orbit. The asymmetrical potential energy profiles can well agree with our design target, as illustrated in Fig. 1(I), which reveals that our optical Brownian ratchet can be accurately generated in the experiment.

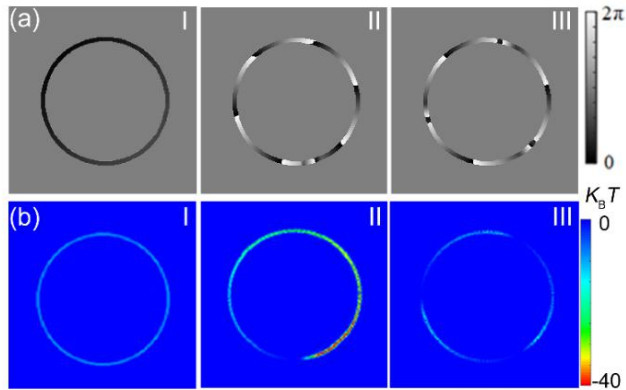

**FIG. 2.** Optical Brownian ratchet by switching ring traps with different potential energy profiles. (a) Phase profiles. (b) Potential energy profiles. Panels I-III correspond to circle traps with flat potential energy profile, one asymmetrical potential energy profile and three asymmetrical potential energy profiles along the circle orbit.

In the experiment, Au nanoparticles with diameter of 200 nm are employed to demonstrate optical Brownian ratchet in holographic optical trapping system (see experiments in Supplementary Material). Here, the probable movement trajectories of a trapped nanoparticle along the circle orbit are discussed in detail. Its angular values varying with time are shown in Fig. 3, whose duration time is about two periods. For Fig. 3(a), there is only one asymmetrical potential well in ring trap, whose duration time $T_1$, $T_2$ and $T_3$ are all equal to 1.5s, and $l_1$ and $l_2$ is about 42.5 μm and 1.5 μm, respectively. For Fig. 3(b), however, there is three asymmetrical potential wells, and duration time $T_1$, $T_2$ and $T_3$ are all equal to 1s, and $l_1$ and $l_2$ is about 13.1 µm and 1.5 µm, respectively. As seen from Fig. 3, there are four types in two periods of motion, as demonstrated in column from I to IV, because there are two probable modes of motion in an asymmetrical potential well in each period, i.e., moving forward or staying in original potential well. Obviously, when the position of particle is above the dotted horizontal line at the moment of $t_A$ and $t_B$ when duration time $T_2$ just ends and $T_3$ just starts, the particle definitely moves forward. Otherwise, it is trapped in the original potential well. These experimental results are in well consistent with our theoretical prediction. In other words, moving forward at the moment means that the particle is in right region of vertical dashed line (illustrated by red color in Fig. 1(II)), while staying in original potential well indicates that the particle is in left region of vertical dashed line. Additionally, the average velocity of forward motion can reach about 28 µm/s and 13 µm/s in Fig. 3(a-b), respectively, which are much faster than the velocity reported previously. Additionally, the snapshots of nanoparticle motion are also presented in Fig. 3, which denote nanoparticle's position at the moment pointed by the arrows (illustrated by red dot in trajectories). It needs to note that its net movement direction is anticlockwise.

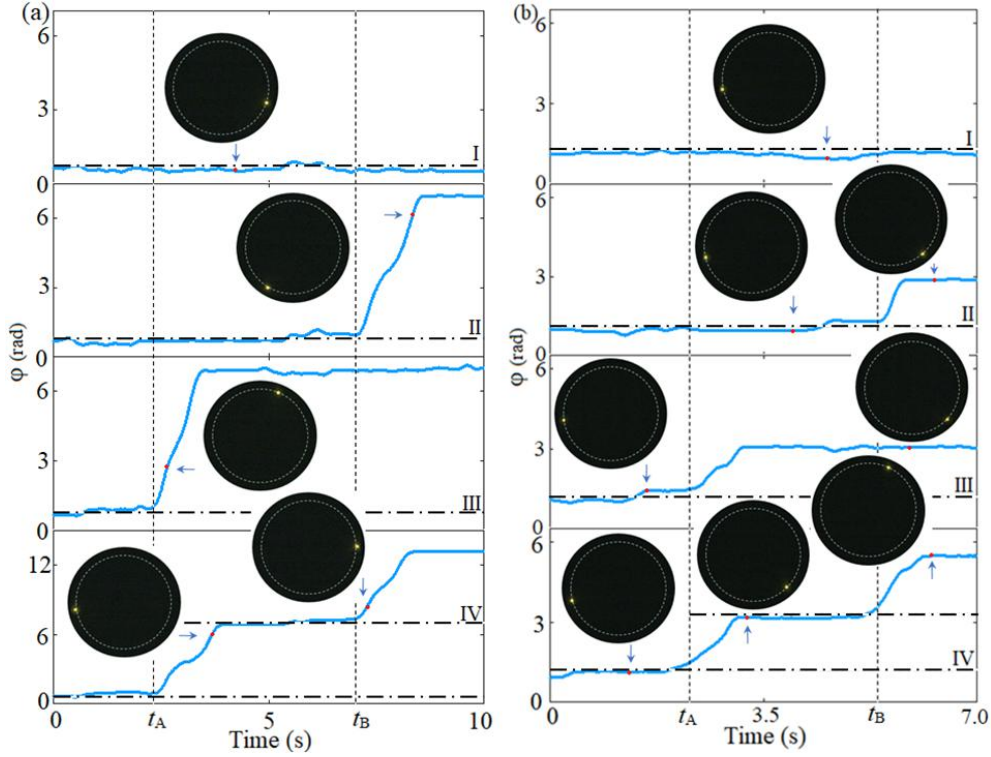

**FIG. 3.** Angular trajectories of nanoparticles as a function of time in optical Brownian ratchet generated by (a) one asymmetrical potential energy profile and (b) three asymmetrical potential energy profiles along the circle orbit. Panels I-IV correspond to four probable motion trajectories in two periods. Note that the nanoparticles rotate in anticlockwise direction.

Furthermore, the position variation of particles in the ring traps with three asymmetrical potential energy profiles in a longer time is also provided, as shown in Fig. 4(a). The duration time $T_1$, $T_2$, $T_3$ are 0.8, 0.5, 0.8 s, respectively, which are shorter than those in Fig. 3(b). The dashed lines in Fig. 4(a) denote the moment that $T_2$ just ends and $T_3$ just starts. It further demonstrates that optical Brownian ratchet can be excellently operated in long operation time, in which it moves forward or stays in original potential well. Meanwhile, its snapshots at different time are presented in Fig. 4(b), which shows its concrete position of particle in Brownian ratchet motion along circle orbit. Similarly, the nanoparticle rotates in anticlockwise direction.

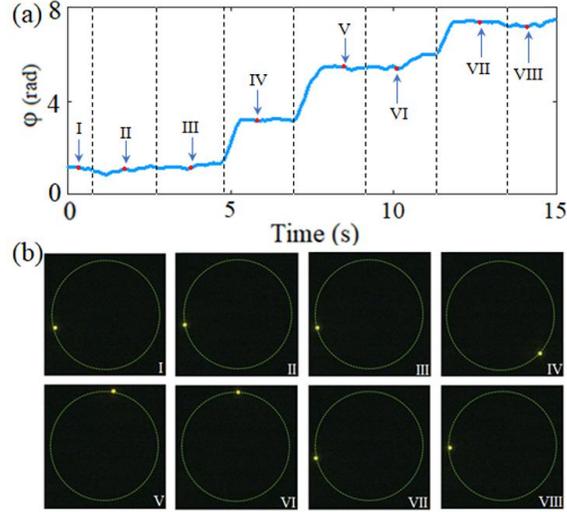

**FIG. 4.** (a) Angular trajectory of nanoparticle in optical Brownian ratchet generated by three asymmetrical potential energy profiles along the circle orbit. (b) Snapshots of nanoparticle motion along the circle orbit. Note that the particles rotate in anticlockwise direction of the circle orbit.

More importantly, optical Brownian ratchet along a rectangle orbit is also demonstrated. It reveals that our optical Brownian ratchet can be easily realized in different motion path, which are highly preferred for potential application in controllable transport of particles. Frankly, although the intensity uniformity could slightly be deteriorated in the experiment, the manipulation capability are not substantially affected, because its potential wells are largely determined by phase gradient. Consequently, accurately creating phase profiles are highly significant in optical Brownian ratchet.

In conclusion, we have proposed optical Brownian ratchet by employing phase profiles in holographic optical trapping system. The dynamical switch between asymmetrical and flat potential energy profiles can be easily manipulated by spatial light modulator (SLM), controlled by Labview software. The experimental results show that average speed of 28 μm/s can be easily obtained, and step distance of about 42 μm can be achieved, for the first time. More importantly, not only its velocity can be variable by changing its phase gradient profile, but motion path and step distance can be also changed by designing different holograms, which are very helpful in optical Brownian ratchets. Consequently, this study provides an opportunity for developing novel functions in transport and manipulation of nanoparticles.

See the supplementary material for Hologram design, experiments and additional results.

# Supplementary Material

# Experimental demonstration of novel optical Brownian ratchet by controllable phase profile of light

Xionggui Tang, Yanhua Xu

*Department of Physics, Key Laboratory of Low Dimensional Quantum Structures and Quantum Control of Ministry of Education, Synergetic Innovation Center for Quantum Effects and Applications, Hunan Normal University, Changsha, 410081, P.R. China*

**METHODS**

**1. Hologram design**

To accurately obtain hologram, a direct method is employed, in which a random phase factor introduced in discrete inverse Fourier transform formula. It is a one-step calculation, whose formula is expressed as follows[1],

$$H(m,n) = c \sum_{\substack{k=-K/2 \\ s=-S/2}}^{\substack{k=K/2 \\ s=S/2}} U(k,s) \exp[j\frac{2\pi}{\lambda f}(km\Delta x_o \Delta x_i + sn\Delta y_o \Delta y_i) + j\phi], \qquad (1)$$

where $\Delta x_o$ and $\Delta y_o$ are the single pixel size in the *x* and *y* direction at the output plane; $\Delta x_i$ and $\Delta y_i$ are the single pixel size in the *x* and *y* direction at the input plane; *j* is a complex unit; $\lambda$ is operation wavelength. *f* is focus length of objective lens; *c* is a constant; $k \in [-K/2, K/2]$, $s \in [-S/2, S/2]$, $m \in [-M/2, M/2]$, $n \in [-N/2, N/2]$, in which *K* and *S* stand for the maximum number of sampling pixels in the *x* and *y* axis at the output plane, respectively, and *M, N* denote the maximum number of sampling pixels in the *x* and *y* axis at the input plane, respectively; $U(k,s)$ stands for $U(k\Delta x_o, s\Delta y_o)$, and $H(m,n)$ denotes $H(m\Delta x_i, n\Delta y_i)$; $U(k,s)$ is target optical field at back focus plane of objective lens; $H(m,n)$ is optical field reflected by SLM at front focus plane of objective lens; $\phi$ is a random phase factor, in which $\phi = \pi \cdot rand$, $rand \in [0,1]$. Consequently, the phase of hologram is written as,

$$P(m,n) = \arg[H(m,n)], \tag{2}$$

where symbol arg stands for phase angle of $H(m,n)$. In this case, we can easily generate holograms once the target optical field is provided.

## 2. Experiments

In the experiment, a CW tunable Ti:Sapphire laser (Spectra-Physics 3900s) with wavelength of 800 nm is operated in a TEM00 Gaussian mode. A phase-only SLM (Hamamatsu X13138) is employed to generate the optical ratchets using holograms designed with a method we developed.[1] An inverted microscope (Olympus IX73) with a 60X objective (NA = 1.2, Olympus UPLSAPO 60XW) is utilized for focusing the laser beam and imaging the nanoparticles. A beam profiler (Edmund Optics) is used for capturing the optical intensity of the reconstructed image. The optical power of laser source is about 191 mW. The Au nanoparticles (NPs) with diameter of 200 nm are purchased from nanoComposix Inc., which are nearly spherical. The trapped Au NPs dispersed in the deionized water, are pushed by the beam against the top interface of the sample cell, and are driven by our optical traps along the 2D orbit. They are visualized by darkfield microscopy (Olympus U-DCW condenser), and recorded at frame rate of 150 fps, by using a CMOS camera (Point-Grey Grasshopper 3).

**FIGURES**

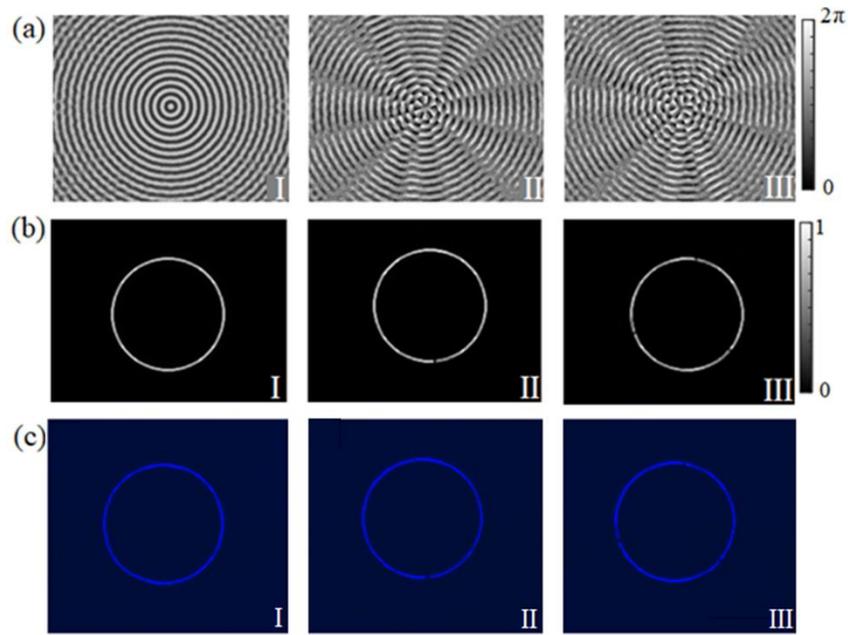

**FIG. S1.** Optical Brownian ratchet by switching ring traps with different potential energy profiles. (a) Holograms. (b) Intensity profiles. (c) Measured intensity distributions. Panels I-III correspond to circle traps with flat potential energy profile, one asymmetrical potential energy profile and three asymmetrical potential energy profiles along circle orbit. Note that the ring radius is about about 7 μm.

**Reference:**

1. X. Tang, F. Nan, F. Han, Z. Yan, Rapidly and accurately shaping the intensity and phase of light for optical nano-manipulation, Nanoscale Advances, 2020, DOI: 10.1039/d0na00167h.